\documentclass[5p,authoryear]{elsarticle}
\usepackage{graphicx}
\usepackage{url}
\usepackage{common}
\usepackage{pdfpages}
\usepackage{times}
\usepackage{helvet}
\usepackage{courier}
\usepackage{subcaption}
\usepackage{url}

\let\B\textbf

\usepackage{array}
\usepackage{ragged2e}
\newcolumntype{P}[1]{>{\centering\arraybackslash}p{#1}}
\journal{}

\makeatletter
\def\@author#1{\g@addto@macro\elsauthors{\normalsize%
    \def\baselinestretch{1}%
    \upshape\authorsep#1\unskip\textsuperscript{%
      \ifx\@fnmark\@empty\else\unskip\sep\@fnmark\let\sep=,\fi
      \ifx\@corref\@empty\else\unskip\sep\@corref\let\sep=,\fi
      }%
    \def\authorsep{\unskip,\space}%
    \global\let\@fnmark\@empty
    \global\let\@corref\@empty  
    \global\let\sep\@empty}%
    \@eadauthor={#1}
}
\makeatother

\begin{document}


\begin{frontmatter}
\title{Stability of Topic Modeling via Matrix Factorization}
\author{Mark Belford\corref{cor1}}
\ead{mark.belford@insight-centre.org}

\author{Brian Mac Namee}
\ead{brian.macnamee@ucd.ie}

\author{Derek Greene}
\ead{derek.greene@ucd.ie}

\address{Insight Centre for Data Analytics, University College Dublin, Ireland}

\begin{abstract}
 
Topic models can provide us with an insight into the underlying latent structure of a large corpus of documents. A range of methods have been proposed in the literature, including probabilistic topic models and techniques based on matrix factorization. However, in both cases, standard implementations rely on stochastic elements in their initialization phase, which can potentially lead to different results being generated on the same corpus when using the same parameter values. This corresponds to the concept of ``instability'' which has previously been studied in the context of $k$-means clustering. In many applications of topic modeling, this problem of instability is not considered and topic models are treated as being definitive, even though the results may change considerably if the initialization process is altered. In this paper we demonstrate the inherent instability of popular topic modeling approaches, using a number of new measures to assess stability. To address this issue in the context of matrix factorization for topic modeling, we propose the use of ensemble learning strategies. Based on experiments performed on annotated text corpora, we show that a K-Fold ensemble strategy, combining both ensembles and structured initialization, can significantly reduce instability, while simultaneously yielding more accurate topic models.
\end{abstract}

\end{frontmatter}

￿
\section{Introduction}
\label{sec:intro}

Topic models aim to discover the latent semantic structure or topics within a corpus of documents, which can be derived from co-occurrences of words across the documents. Popular approaches for topic modeling have involved the application of probabilistic algorithms such as Latent Dirichlet Allocation \citep{blei03lda}. More recently, Non-negative Matrix Factorization approaches \citep{lee99nmf} have also been successfully applied to identify topics in unstructured text \citep{arora12beyond,kuang15nmfbook}. 

The standard formulations of both the LDA and NMF algorithms include stochastic elements in their initialization phase, prior to an optimization phase which produces a local solution. This random component can affect the final composition of the topics found and the rankings of the terms that describe those topics. This is problematic when seeking to capture a definitive topic modeling solution for a given corpus and represents a fundamental \emph{instability} in these algorithms -- different runs of the same algorithm on the same data can produce different outcomes. This problem has been widely studied in the context of partitional clustering algorithms such as $k$-means, which tends to converge to one of numerous local minima, depending on the choice of starting condition \citep{bradley98refining}. It has long been recognized as a significant drawback of such algorithms and a substantial number of works exist which attempt to address the issue (\eg \cite{pena99init},\cite{kuncheva06evaluation}).

In the case of topic modeling, instability can manifest itself in two distinct aspects. The first can be observed when examining the topic descriptors (\ie the top terms representing each topic) over multiple runs. The term rankings may change considerably, where certain terms may appear or disappear completely between runs. 
Secondly, issues of instability can also be observed when examining the degree to which documents have been associated with topics across different runs of the same algorithm on the same corpus. In both cases, such inconsistencies can potentially alter our interpretation and perception of a given topic model. Also, it is clear that any individual run should not be treated as a ``definitive'' summary of the underlying topics present in the data.

Generally speaking, in the comparative evaluation of topic modeling approaches, researchers tend to focus on either the coherence of the topic descriptors \citep{newman10eval} or the extent to which the topics accurately coincide with a set of ground truth categories or human annotations \citep{kuang15nmfbook}. However, few researchers have considered the evaluation of different approaches from the point of view of their stability across multiple runs.

In this paper we quantitatively assess the extent to which standard randomly-initialized NMF and LDA algorithms are unstable with respect to the topics that they produce on a diverse collection of text corpora. To do this we propose measures that capture the two distinct aspects of instability outlined above. We then focus on addressing the issue in the context of matrix factorization, exploring the use of strategies that involve improved initialization and ensemble learning. In particular, we propose a new combined approach, motivated by the traditional concept of $k$-fold cross-validation, which can yield stable results while also often producing more accurate and coherent models\footnote{See \url{https://github.com/derekgreene/topic-ensemble}}.

The rest of the paper is structured as follows. In \refsec{sec:related} we provide an overview of relevant work in topic modeling and the more general area of cluster analysis. In \refsec{sec:stability} we discuss the problem of topic model instability in more detail, describing three new measures to quantify instability in topic models. In \refsec{sec:methods} we propose ensemble approaches to address the issue, which are subsequently evaluated on ten different text corpora in \refsec{sec:eval}. Finally in \refsec{sec:conc} we conclude the paper with ideas for future work.

￿
\section{Related Work}
\label{sec:related}

\subsection{Topic Modeling}  

Topic models attempt to discover the hidden thematic structure within an unstructured collection of text without relying on any form of training data. These models date back to the early work on latent semantic analysis (LSA) by \cite{deerwester90lsi}, who proposed applying SVD to decompose a document-term matrix to uncover the associations between terms and concepts in the data. In basic terms, a topic model consists of $k$ topics, each represented by a ranked list of strongly-associated terms (often referred to as a ``topic descriptor''). Each document in the corpus can also be associated with one or more of these topics to varying degrees. 

Considerable research on topic modeling has focused on the use of probabilistic methods, where a topic is viewed as a probability distribution over words, with documents being mixtures of topics \citep{steyvers06prob}. The most widely-applied probabilistic topic modeling approach has been LDA \citep{blei03lda}. Different approximation methods have been proposed for LDA inference, including variational inference and Markov chain Monte Carlo (MCMC). Such approximation algorithms can converge to different local maxima on the same data \citep{zhao15heuristic}. The most commonly-used implementation, provided by the \emph{Mallet} software package \citep{mccallum02mallet}, relies on fast Gibbs sampling, where the initial state is determined by a user-specified random seed. 

Alternative algorithms, such as Non-negative Matrix Factorization \citep{lee99nmf}, have also been effective in discovering topics in text corpora \citep{arora12beyond,kuang15nmfbook}.  NMF is an unsupervised approach for reducing the dimensionality of non-negative matrices. When working with a document-term matrix $\m{A}$, the goal of NMF is to approximate this matrix as the product of two non-negative factors $\m{W}$ and $\m{H}$, each with $k$ dimensions. The rows of the factor $\m{H}$ can be interpreted as $k$ topics, defined by non-negative weights for each of the $m$ terms in the corpus vocabulary. Ordering each row provides a topic descriptor, in the form of a ranking of the terms relative to the corresponding topic. The columns in the matrix $\m{W}$ provide membership weights for all documents with respect to each of the $k$ topics.
One of the advantages of NMF over traditional LDA methods is that there are fewer parameter choices involved in the modeling process, while it also has a tendency to identify more coherent topics than LDA \citep{ocallaghan15eswa}. 

NMF is commonly initialized by assigning random non-negative weights to the entries in the factors $\m{W}$ and $\m{H}$. By applying an optimization process, such as alternating least squares \citep{lin07nmf}, the factors are iteratively improved to reduce the approximation error until a local minimum is reached. As a result, the values in the initial pair of factors will have a significant impact on the values in the final factors (\ie the topic-term and topic-document weights), even after a large number of iterations have been performed. Alternative initialization schemes for NMF have focused on increasing the accuracy of the final factors by using a more structured process, such as seeding using a prior clustering algorithm \citep{wild04init}. Another approach, Non-negative Double Singular Value Decomposition (NNDSVD) \citep{bout08headstart}, chooses initial factors based on a sparse SVD approximation of the original data matrix. This has been shown to be particularly effective on sparse data, such as text \citep{ocallaghan15eswa}. In its basic form, NNDSVD contains no stochastic element and should technically converge to the same pair of factors each time, although this depends on the underlying SVD implementation being used.

\subsection{Stability in Cluster Analysis}

Partitional clustering algorithms, such as $k$-means and $k$-medoids, have an inherent stability problem. That is to say, if we run the same algorithm on the same data or data drawn from the same source repeatedly, we frequently achieve different results between each run. This variation can either be due to poor random seeds leading to convergence to different local minima \citep{pena99init}, or as a result of perturbations in the data \citep{benhur02stability}.

One widely-adopted approach for dealing with the issue is to adopt a better cluster initialization strategy that is either fully deterministic or at least produces less variation than random initialization, while simultaneously yielding more useful clusterings. 
A popular initialization approach proposed by \cite{arthur07kpp}, referred to as \emph{$k$-means$++$}, involves choosing an initial seed item at random as the first cluster center and then choosing each subsequent cluster center with a probability proportional to its squared distance from the items nearest existing cluster centers. To further improve the resulting clustering, this process can be repeated for several different initial seed items. While this strategy is not deterministic, it does tend to yield more consistent results across multiple runs. Researchers have also proposed fully deterministic strategies, where initial cluster centers are determined based on embedding methods such as PCA \citep{su04deterministic}, or coming from the prior application of another algorithm such as hierarchical clustering \citep{celebi12deterministic}.

\subsection{Ensemble Methods}

An alternative strategy for reducing instability in unsupervised learning is to use ensemble clustering techniques, which are based on the premise that combining large, diverse sets of clusterings can produce a more stable and accurate solution \citep{strehl02multiple}. Ensemble approaches are usually divided into two different stages. Firstly, a collection of base clusterings are generated (\ie the ensemble members), typically by repeatedly applying an algorithm such as $k$-means with random initialization to the full dataset or to random samples of the data \citep{bidgoli04resampling}. Secondly, an integration function is applied to combine the base clusterings into a single consensus clustering. One of the most common integration strategies utilized is to leverage information from the ensemble regarding the level of ``co-association'' between all pairs of items. The underlying idea behind this is that items that are frequently assigned together in different clusterings will naturally belong to the same underlying group \citep{strehl02multiple}. The resulting consensus clustering represents an approximation of the ``average'' clustering from among the ensemble members. \citet{hadjitodorov06moderate} demonstrated a trade-off between diversity and quality in cluster ensembles, and proposed a number of measures to quantify diversity for such ensembles.

Work regarding the optimality and consistency of solutions produced by clustering and biclustering algorithms has been previously carried out in other domains \citep{bertoni05stability,pio2015comirnet}. However, in the general area of matrix factorization, there has been only some initial work on the use of ensemble approaches. This includes using a hierarchical scheme to combine multiple factorizations in the study of protein networks \citep{greene08ensemble} and the generation of ensembles of factorizations via a boosting-like approach \citep{suh16ensnmf}. Recent work has also looked at using stability as a means of identifying an appropriate number of topics in a given corpus when applying NMF to text data \citep{greene14topics}. However, these studies have not investigated the extent of the problems introduced by instability in the context of topic modeling. 
￿
\section{Stability of Topic Modeling}
\label{sec:stability}

In this section we introduce three new measures for assessing the stability of a collection of topic models, and use these to demonstrate how standard NMF and LDA approaches can be prone to produce unstable results when applied to text corpora.

\subsection{Overview}

As discussed in \refsec{sec:related}, standard implementations of topic modeling approaches, such as LDA and NMF, commonly employ stochastic initialization prior to optimization. As a result, the models they produce can vary quite considerably between different runs. Regardless of whether we are applying probabilistic or non-probabilistic algorithms, we can observe that this variation manifests itself in two ways: in relation to term-topic associations, or document-topic associations. In the former, the ranking of the top terms that describe a topic can change significantly between runs. In the latter, documents may be strongly associated with a given topic in one run, but may be more closely associated with an alternative topic in another run. In more extreme cases, a consequence of both manifestations is that topics can ``appear'' or ``disappear'' across different runs of the algorithm. This presents a challenge for domain experts who seek to gain a reliable insight into a particular corpus of documents. Depending on the topics resulting from a given algorithm run, their interpretation of the data may change considerably. However, the implications of this variation are rarely discussed in the topic modeling literature, particularly in the context of matrix factorization. 

\subsection{Measuring Stability}
\label{sec:measures}

To actually quantify the level of stability/instability present in a collection of topic models $\fullset{\aset{M}}{r}$ generated over $r$ runs on the same corpus, we propose three measures which reflect both aspects of topic model stability as described above. These measures are general in the sense that they can be applied to models generated using either probabilistic or matrix factorization algorithms.

\subsubsection{Descriptor Set Difference}
If the topics present in two topic models are similar, we should naturally expect that the prominent terms appearing in the topic descriptors in both models will be similar. Formally, if we represent each topic in a single model $\aset{M}_{i}$ with a number of top-ranked terms $t$, we can calculate the \emph{descriptor set} as the union of top terms across all $k$ topics, which we denote $T_{i}$. By measuring the symmetric difference between the descriptor sets for two different models, we can broadly gauge the similarity of the two models. This is useful as we can capture the variance at the descriptor level as terms may appear and disappear between runs. Formally, given two topic models $\aset{M}_{i}$ and $\aset{M}_{j}$, each containing $k$ topics represented by their top $t$ terms, we calculate the \emph{descriptor set difference} as:

\begin{equation}
DSD( \aset{M}_{i},\aset{M}_{j}) = \frac{ T_{i} \triangle T_{j} }{t \times k}  
\label{eqn:dsd}
\end{equation}
A value of 0 indicates identical descriptor sets (\ie no difference), while a value of 1 indicates that the topic descriptors for the two models share no common terms at all. Given a collection of $r$ topic models, we can calculate the Average Descriptor Set Difference (ADSD):
\begin{equation}
ADSD = \frac{1}{r \times (r-1)} \sum_{i,j,i\neq j}^{r} DSD( \aset{M}_{i},\aset{M}_{j}) 
\label{eqn:adsd}
\end{equation}
This produces a value $\in [0,1]$, where a value closer to 0 for \reft{eqn:adsd} is indicative of a more stable collection of models.
	
\subsubsection{Topic-Term Stability}
While the ADSD gives an overall measure of the difference between two models, it does not account for cases where topics are ``mixed'' across different runs of the algorithm (\ie the same terms appear in different topics across different runs). Therefore, we propose a measure that compares the similarity between two topic models based on a pairwise matching process at the topic level. This is important as topics may appear and disappear between different runs and also helps to capture the variance at the individual topic level.

First, given a pair of individual topics represented by their top $t$ terms, we can measure the similarity between them based on the Jaccard Index:
\begin{equation}
\label{eqn:jac}
Jac(R_{i},R_{j}) = \frac{\abs{ R_{i} \cap R_{j} } }{ \abs{ R_{i} \cup R_{j} } }
\end{equation}
where $R_{i}$ denotes the top $t$ ranked terms for the $i$-th topic (\ie its topic descriptor). We can use the above to build a measure of the agreement between two complete topic models, each containing $k$ topics. We construct a $k \times k$ similarity matrix $\m{S}$, such that the entry $S_{xy}$ indicates the agreement between the $x$-th topic in the first model and the $y$-th topic in the second model, as calculated using \reft{eqn:jac}.  We then find the best match between the rows and columns of $\m{S}$ (\ie the topics from the first model and the second model). The optimal permutation $\pi$ may be found in $O(k^{3})$ time by solving the minimal weight bipartite matching problem using the Hungarian method \citep{kuhn55hungarian}. From this, we can produce a Term Stability (TS) score:
\begin{equation}
TS(\aset{M}_{i},\aset{M}_{j}) = \frac{1}{k}\sum_{x=1}^{k} Jac(R_{ix},\pi(R_{ix}))
\label{eqn:agree}
\end{equation}
where $\pi(R_{ix})$ denotes the topic in model $\aset{M}_{j}$ matched to $R_{ix}$ in model $\aset{M}_{i}$ by the permutation $\pi$. Values for the above take the range $[0,1]$, where a comparison between two identical $k$-way topic models will result in a score of 1.

For a collection of $r$ topic models $\fullset{\aset{M}}{r}$, we can calculate the Average Term Stability (ATS):
\begin{equation}
ATS = \frac{1}{r \times (r-1)} \sum_{i,j,i\neq j}^{r} TS( \aset{M}_{i},\aset{M}_{j}) 
\label{eqn:ats}
\end{equation}
where a score of 1 indicates that all pairs of topic descriptors matched together across the $r$ runs contain the same top $t$ terms.

\subsubsection{Partition Stability}

The second manifestation of topic model instability relates to document-topic associations. To measure the extent to which the associations between a document and one or more topics varies across different runs, for each run we can look at the dominant topic for every document. That is, we convert the document-topic associations (the probabilities in the case of LDA or the $\m{W}$ factor weights in the case of NMF) into a disjoint partition by taking the maximum value for each document. We can then compare the similarity between the partitions generated in two runs using standard clustering agreement measures. Utilizing this information allows us to observe if the dominant topic for each document changes frequently between runs. One widely-used such measure is Normalized Mutual Information (NMI) \citep{strehl02multiple}, which quantifies the level of agreement between two partitions $P_{i}$ and $P_{j}$: 
\begin{equation}
NMI(P_{i}, P_{j}) = \frac{I(P_{i},P_{j})}{\sqrt{H(P_{i})H(P_{j})}} 
\label{eqn:nmi}
\end{equation}
where $I(P_{i},P_{j})$ is the mutual information between the assignments in the two partitions and $H(P_{i})$ is the entropy of the assignments in $P_{i}$ alone. 

We can compute the overall level of agreement between a set of $r$ partitions generated by $r$ runs of an algorithm on the same corpus as the mean Pairwise Normalized Mutual Information (PNMI) for all pairs:
\begin{equation}
PNMI = \frac{1}{r \times (r-1)} \sum_{i,j,i\neq j}^{r} NMI( P_{i},P_{j}) 
\label{eqn:pnmi}
\end{equation}
where $P_{i}$ is the partition produced from the document-topic associations in model $\aset{M}_{i}$. If the partitions across all models are identical, PNMI will yield a value of 1.

\subsection{Examples of Instability}
\label{sec:examples}

\begin{figure}[!b]
    \centering
    \includegraphics[width=\linewidth]{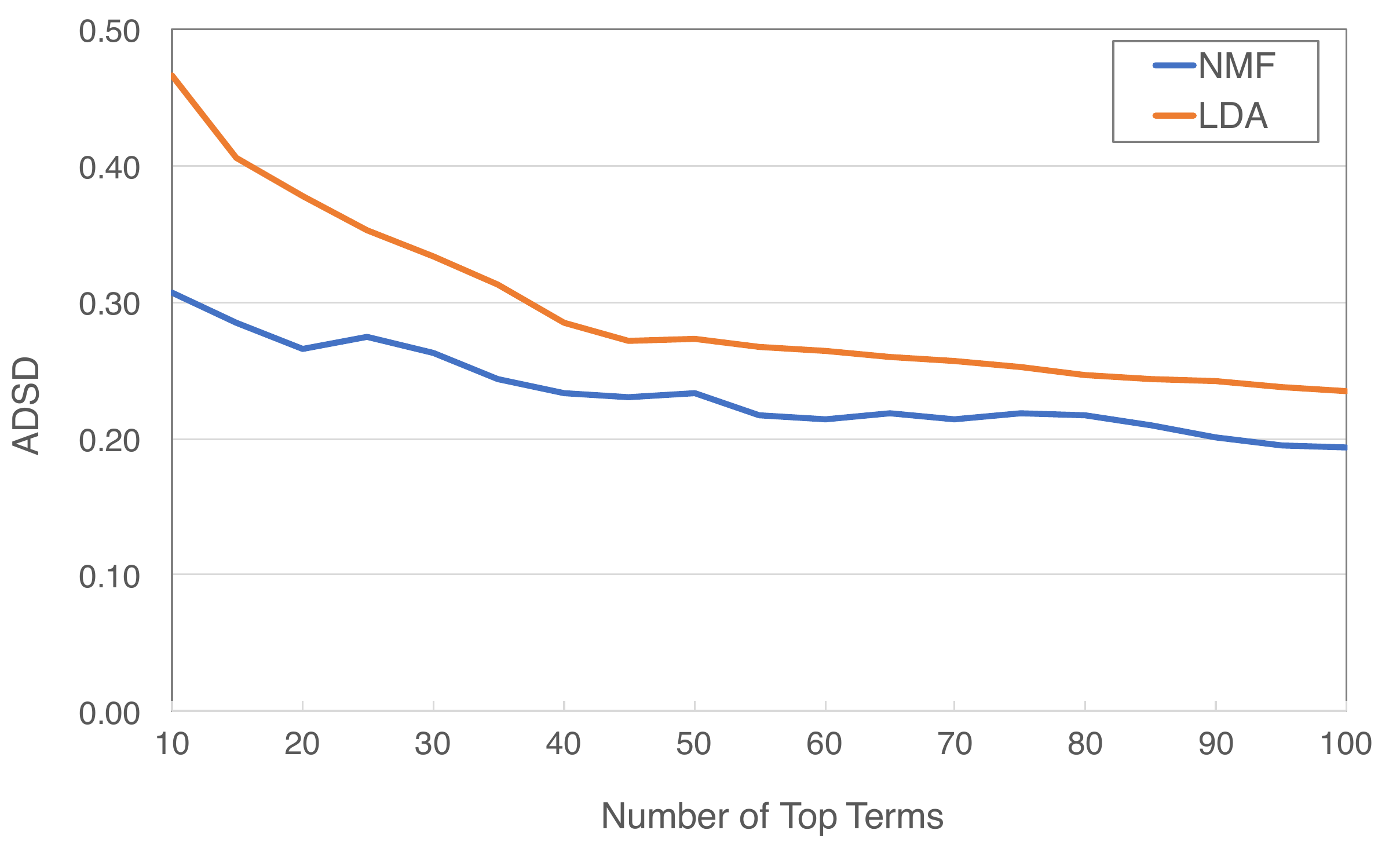}
	\caption{Plot of the Average Descriptor Set Difference (ADSD) for 100 runs of NMF and LDA on the \emph{nytimes-2003} corpus, for increasing numbers of top terms representing each topic.}
    \label{fig:instability}
\end{figure}
\begin{figure}[!b]
\centering
   \begin{subfigure}[b]{0.48\textwidth}
   \includegraphics[width=1\linewidth]{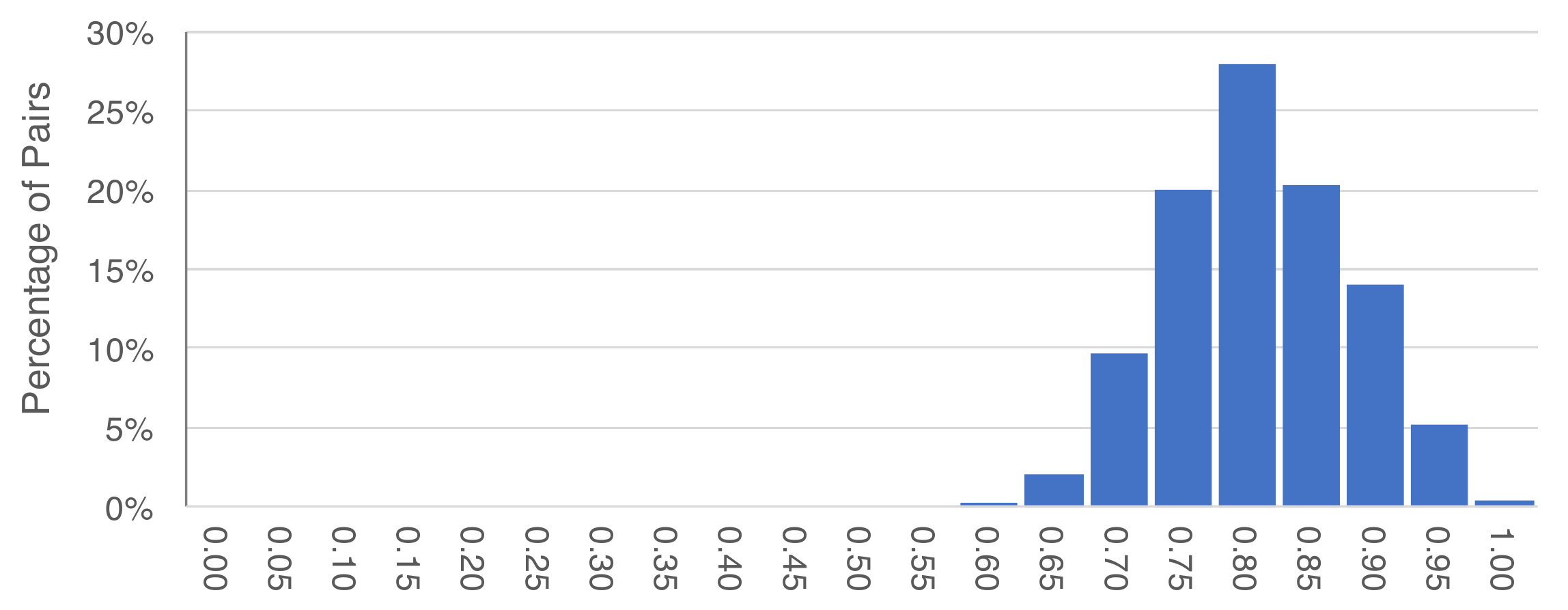}
   \caption{NMF}
\end{subfigure}
\begin{subfigure}[b]{0.48\textwidth}
   \includegraphics[width=1\linewidth]{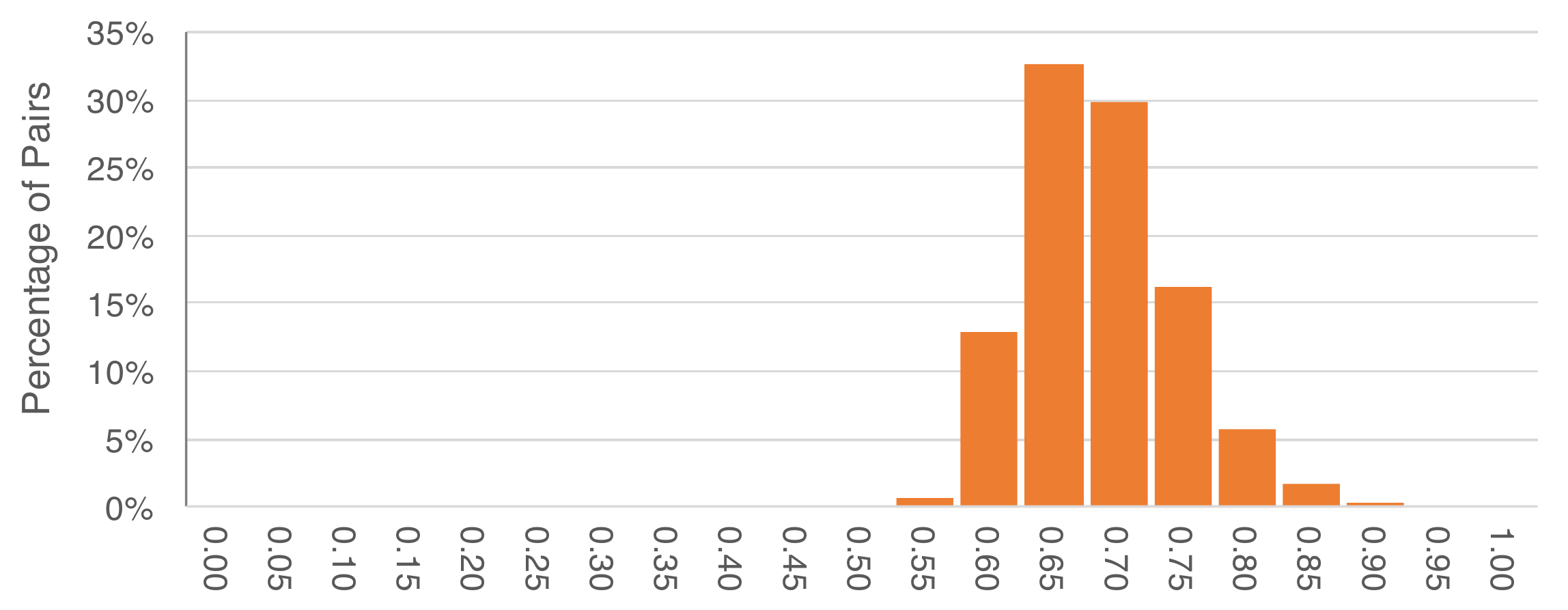}
   \caption{LDA}
\end{subfigure}
\caption{Distribution of pairwise NMI agreement scores for document partitions resulting from 100 runs of (a) NMF, (b) LDA on the \emph{nytimes-2003} corpus.}
\label{fig:docinstability}
\end{figure}

We now provide examples of how the measures proposed above can reflect the problem of instability, using a corpus of news articles from the New York Times as an example (the \emph{nytimes-2003} corpus described later in \refsec{sec:eval}). 

Firstly, to illustrate the issue of term instability, we consider topic models generated for $r=100$ runs of randomly-initialized NMF and LDA, with a fixed number of topics $k=7$ (corresponding to the number of annotated categories in the data). \reffig{fig:instability} shows the ADSD scores for each algorithm, as the number of top terms in the topic descriptors increases from 10 to 100. We can observe that, even with this relatively relaxed measure, there exists substantial variation in the terms appearing in the models for both algorithms. If we represent each topic using 10 terms, the ADSD score is as high as $0.47$ for LDA and $0.31$ for NMF. Even when we extend the topic descriptors to contain 100 terms, which we might expect to capture the bulk of the key terms for the topics in this corpus, the mean difference across runs is $0.23$ and $0.19$ respectively.

We can explore this term instability further by inspecting the topic-term stability of the topics. As an example, \reftab{tab:stability} refers to five separate runs of a related topic (corresponding to the category ``sport'' in the ground truth  for this corpus) for NMF and LDA. For both algorithms it is clear that the ordering of the top terms for this topic can change considerably, and it is also possible for terms to completely disappear between different runs.

￿
\begin{table}[!t]
\begin{center}
\caption{Top-ranked terms for topics related to ``sport'', generated across five runs of randomly-initialized NMF and LDA applied to a news corpus.}
\vskip 0.5em
\begin{scriptsize}
\begin{tabular}{c|p{3.15in}}
\hline
\textbf{\#} & \textbf{LDA Top 10 Terms}                                                                      \\ \hline
1                     & game, team, season, play, coach, games, points, players, against, football  \\
2                     & game, season, team, coach, play, games points, league, football, players \\
3                     & game, season, coach, team, football, league, giants, play, jets, players  \\
4                     & game, season, team, yankees, games, play, mets, nets, left, league  \\
5                     & game, team, season, play, games, players, coach, yankees, time, against        \\ \hline
\textbf{\#} & \textbf{NMF Top 10 Terms}                                                                      \\ \hline
1                     & game, season, team, yankees, games, nets, play, points, players, coach     \\
2                     & game, team, season, nets, points, games, coach, play, knicks, players \\
3                     & game, team, season, nets, points, games, play, coach, knicks, giants     \\
4                     & game, nets, team, season, coach, points, knicks, jets, giants, play      \\
5                     & game, team, season, nets, points, games, play, knicks, coach, kidd \\ \hline
\end{tabular}
\end{scriptsize}
\label{tab:stability}
\end{center}
\end{table}

Using the same corpus, we can also explore the partition stability afforded by both NMF and LDA. \reffig{fig:docinstability} plots the distributions for the NMI agreement scores between all pairs of partitions corresponding to the set of 100 models produced by each algorithm. Two partitions with identical document assignments would yield an NMI score of 1. However, only $0.4\%$ of all pairs of partitions achieve this for NMF. In the case of LDA, of the 4950 unique pairs of partitions, none achieve perfect agreement and only $0.3\%$ achieve an NMI score $\geq 0.9$. When we average the agreement scores over all runs, the overall PNMI scores for the two algorithms are 0.78 and 0.66 respectively, indicating there is considerable variation in the outputs of both algorithms across these runs. 

Again, manually inspecting the top-ranked documents in the topics for these models reveals the extent of the variation. \reftab{tab:document-stability-example} lists the identifiers of the top ten documents assigned to each of the topics related to ``sport'' selected from five runs of NMF and LDA in \reftab{tab:stability}. We observe that, similar to the case of the top terms, the ordering of documents is also subject to the same inherent instability.

\begin{table}[!t]
\begin{center}
\caption{Top-ranked documents for topics related to ``sport'', generated across five runs of randomly-initialized NMF and LDA applied to a news corpus.}
\vskip 0.5em
\begin{scriptsize}
\begin{tabular}{c|p{3.15in}}
\hline

\textbf{\#} & \textbf{LDA Top 10 Documents}                                                                      \\ \hline
1                     & s4310, s5376, s4247, s5262, s6055, s1493, s3167, s5670, s4972, s6636     \\ 
2                     & s4441, s6267, s4247, s3521, s8146, s5262, s4708, s4681, s4460, s8937  \\ 
3                     & s0113, s3521, s4708, s1698, s5299, s4972, s8577, s8351, s5855, s6834     \\ 
4                     & s6267, s0113, s5376, s9894, s3521, s5262, s9116, s4681, s4708, s8937       \\ 
5                     & s6267, s0113, s9894, s4247, s5262, s4681, s2056, s4972, s8577, s6636     \\ \hline
\textbf{\#} & \textbf{NMF Top 10 Documents}                                                                      \\ \hline
1                     & s5995, s6558, s3547, s9993, s5281, s8029, s2484, s5114, s1227, s2934     \\ 
2                     & s6558, s5995, s8029, s5457, s5843, s2484, s9993, s1227, s5193, s2068 \\ 
3                     & s8029, s6558, s5995, s5457, s5843, s2484, s1227, s9993, s5193, s2068     \\ 
4                     & s5457, s8029, s5843, s6558, s9993, s5193, s2068, s7687, s2484, s9924      \\ 
5                     & s8029, s6558, s5995, s5457, s5843, s2484, s1227, s9993, s5193, s2265\\ \hline
\end{tabular}
\end{scriptsize}
\label{tab:document-stability-example}
\end{center}
\end{table}


\subsection{Redundant Stability}
\label{sec:redundant}

While the examples above demonstrate that the production of robust, reliable topic models is important, it is also necessary to emphasize that stability should not be the sole requirement for a useful topic modeling algorithm. As observed by \citet{benhur02stability} in the context of partitional clustering, in some situations stability can simply be indicative of an algorithm's tendency to converge to a given local solution, regardless of the quality of that solution. In the context of NMF, we could initialize the factors $\m{W}$ and $\m{H}$ in a deterministic way with arbitrary non-negative values. However, this ``redundant stability'' is unlikely to provide a useful model. Therefore, in the next section we propose techniques that yield solutions that are not only stable but also accurate -- \ie the topics are semantically coherent and provide a useful insight into the content of the corpus.

￿
\begin{figure*}[!th]
    \centering
    \includegraphics[width=0.8\linewidth]{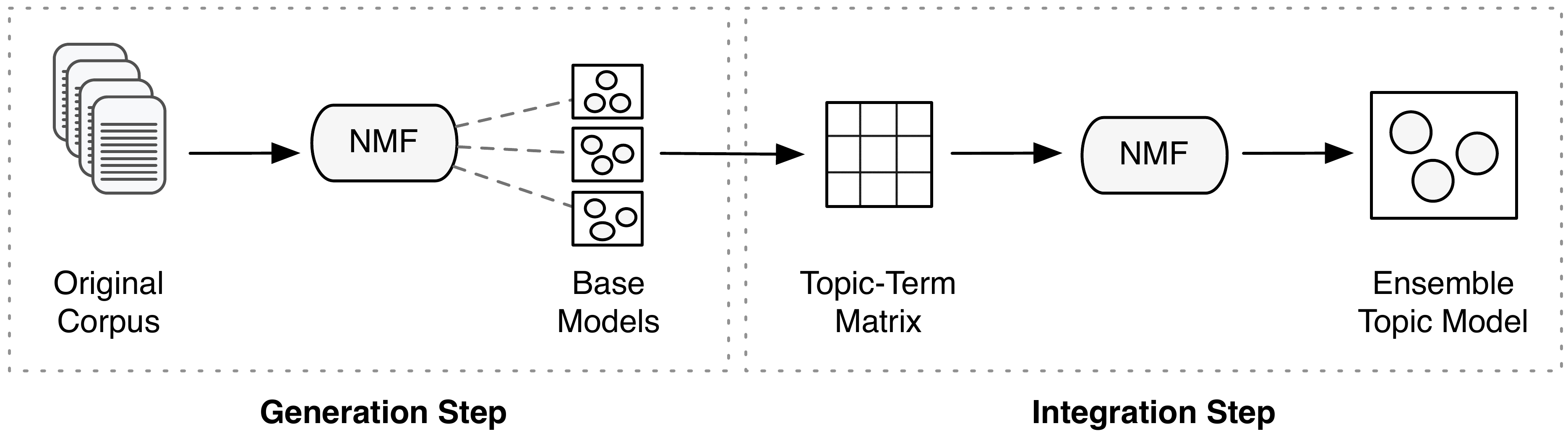}
    \vskip -0.2em
	\caption{Overview of a general ensemble strategy for topic modeling with matrix factorization, consisting of two steps: generation and integration.}
    \label{fig:ensemble}
\end{figure*}

\section{Methods}
\label{sec:methods}

We now propose ensemble methods for topic modeling via matrix factorization, which can be utilized to address the issue of stability, while also potentially producing more accurate topic models for a corpus of unstructured text.

\subsection{Basic Ensemble Method}
\label{sec:ensemble}

We apply ensemble learning for topic modeling in the form of two layers of matrix factorization. \reffig{fig:ensemble} shows an overview of the method, which can naturally be divided into two steps, similar to existing strategies in ensemble clustering \citep{strehl02multiple}:

\begin{enumerate}
\item \emph{Generation}: Create a set of \emph{base topic models} by executing $r$ runs of NMF applied to the same corpus, represented as a document-term matrix $\m{A}$.
\item \emph{Integration}: Transform the base topic models to a suitable intermediate representation, and apply a final run of NMF to produce a single \emph{ensemble topic model}, which represents the final output of the method.
\end{enumerate}
We now discuss both of these steps in more detail.

\subsubsection{Ensemble Generation}
\label{sec:generation}

Unsupervised ensemble procedures typically seek to encourage diversity with a view to improving the quality of the information available in the integration phase \citep{topchy05weak}. Therefore, we create a diverse set of \textit{r} base topic models (\ie the topic term descriptors and document assignments will differ from one base model to another). Here we encourage diversity by relying on the inherent instability of NMF with random initialization -- we generate each base model by populating the factors $\m{W}$ and $\m{H}$ with values based on a different random seed, and then applying NMF to $\m{A}$. In each case we use a fixed pre-specified value for the number of topics $k$. After each run, the $\m{H}$ factor from the base topic model (\ie the topic-term weight matrix) is stored for later use. 

\subsubsection{Ensemble Integration}
\label{sec:integration}

\begin{figure}[!b]
    \centering
    \includegraphics[width=0.84\linewidth]{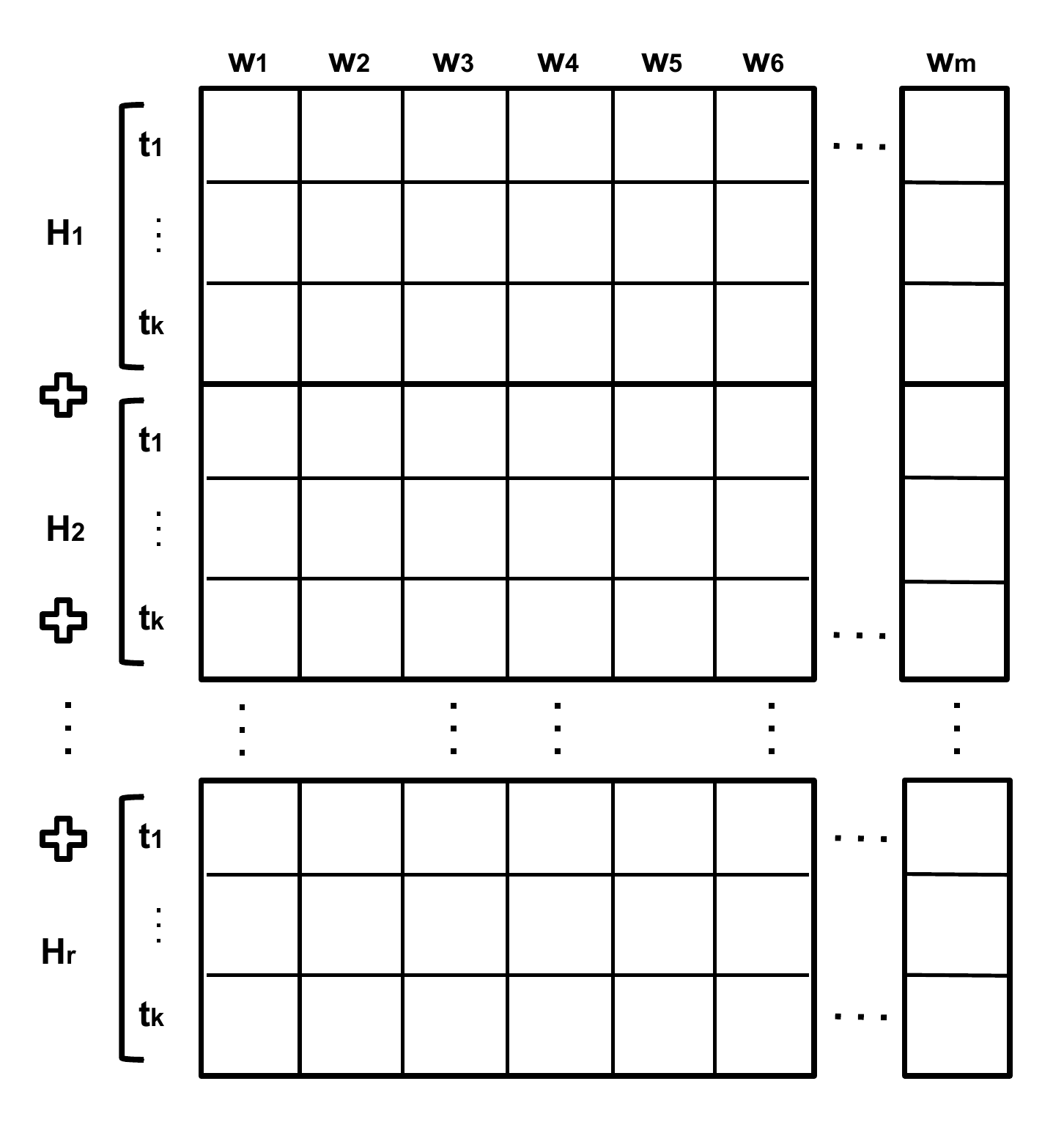}
	\vskip -0.2em
	\caption{Illustration of the stacking process where the $\m{H}$ factors from \emph{r} topic models, each consisting of \emph{k} topics and \emph{m} terms, are combined to create a single topic-term matrix. }
    \label{fig:stacking}
\end{figure}

Once we have generated a collection of $r$ factorizations, in the second step we create a new representation of our corpus in the form of a topic-term matrix $\m{M}$. The matrix is created by stacking the transpose of each $\m{H}$ factor generated in the first step, as illustrated in \reffig{fig:stacking}. Here each factor consists of $k$ topics $\fullset{t}{k}$ and $m$ terms $\fullset{w}{m}$. We construct this topic-term matrix as we may often expect to see similar topics appearing between different runs. However, they may not be identical with respect to their terms, and we wish to leverage this variance. It is important to note that this process of combining the factors is order independent. This results in a matrix $\m{M}$ where each row corresponds to a topic from one of the base topic models, and each column is a term from the original corpus. Each entry $M_{ij}$ holds the weight of association for term $i$ in relation to a single topic from a base model.

Once we have created $\m{M}$, we apply the second layer of NMF to this matrix to produce the final ensemble topic model. The reasoning behind applying NMF a second time to these topic descriptors is that they explicitly capture the variance between the base topic models. To improve the quality of the resulting topics, we generate initial factors using NNDSVD initialization \citep{bout08headstart}. As an input parameter to NMF, we specify a final number of $k'$ topics, which is typically set to be the same as the value $k$ used in the generation step. The resulting $\m{H}$ factor provides weights for the terms for each of the $k'$ ensemble topics -- the top-ranked terms in each column can be used as descriptors for a topic. To produce weights for the original documents in our corpus, we can ``fold'' the documents into the ensemble model by applying a projection to the document-term matrix $\m{A}$:
\[
\m{D} = \m{A} \cdot \mtr{H}
\]
Each row of $\m{D}$ now corresponds to a document, with columns corresponding to the $k'$ ensemble topics. An entry $D_{ij}$ indicates the strength of association of document $i$ in ensemble topic $j$.

\begin{figure}[!t]
\begin{algor}
	\item Construct full document-term matrix $\m{A}$ for the corpus.
	\item For $p$ rounds:
		\begin{enumerate}
		\item Randomly divide corpus into $f$ folds.
		\item For each of the $f$ folds:
			\begin{enumerate}
			\item Exclude the current fold.
			\item Apply NMF with NNDSVD initialization to documents in  $\m{A}$  from the other $(f-1)$ folds to generate $k$ topics.
			\end{enumerate}
		\end{enumerate}
	\item From the resulting $p \times f$ models, construct a topic-term matrix $\m{M}$ by stacking the transpose of each $\m{H}$ factor.
	\item Apply NMF with NNDSVD initialization to $\m{M}$ to generate $k'$ topics, where the factor $\m{H}$ gives the final topic-term associations.
	\item Compute $\m{D} = \m{A} \cdot \mtr{H}$ to find the final document-term associations.
\end{algor}
\caption{Summary of K-Fold ensemble topic modeling method.}
\label{fig:kfold}
\end{figure}

\subsection{K-Fold Ensemble Method}
\label{sec:kfold}
While the basic ensemble generation approach described in \refsec{sec:generation} does yield a diverse set of base topic models, the use of random initialization means that some of these models will correspond to poor local minima with low accuracy. Furthermore, given the number of possible initial factors that could be generated in this way, there is still potential for several runs of the complete ensemble process to yield somewhat different final results. Therefore, we consider the use of improved initialization to generate more accurate base models, while also using a more structured strategy to create the models in order to reduce variability. This strategy is based on traditional k-fold cross-validation as performed in evaluation in supervised learning. 

In our case, we randomly divide the corpus of documents into $f$ folds of equal size. Each of the $f$ folds is excluded in turn, and we apply NMF with NNDSVD initialization to the documents from the remaining $(f-1)$ folds, yielding $f$ models. To reduce variability, we repeat the process for $p$ rounds using different splits of the data, yielding a total of $p \times f$ topic models, each generated on a large subsample of the corpus. This collection of base topic models is then integrated as described in \refsec{sec:integration} to produce a final topic model. A full summary of this approach is given in \reffig{fig:kfold}.

￿
\section{Evaluation}
\label{sec:eval}

￿
\begin{table*}[!h]
	\begin{center}
    \caption{Details of the ten corpora used in our experiments, including the total number of documents $n$, number of terms $m$, and number of categories $\hat{k}$ in the associated ``ground truth'' annotations.}
\vskip -0.5em
    \begin{tabular}{|l|r|r|r|p{11.7cm}|}
    \hline
    \textbf{Corpus}             & \textbf{$n$}\; & \textbf{$m$}\; & \textbf{$\hat{k}$}\;\ & \textbf{Description} \\
    \hline
    bbc                 & 2,225\;     & 3,121\;     & 5\;  & General news articles from the BBC from 2003. \\
   bbc-sport            & 737\;      & 969\;      & 5\; & Sports news articles from the BBC from 2003. \\
    guardian-2013   	& 6,520\;     & 10,801\;    & 6\;  &  Corpus of news articles published by The Guardian during 2013. \\
   irishtimes-2013    & 3,246\;     & 4,832\;     & 7\;  & Corpus of news articles published by The Irish Times during 2013.\\
	nytimes-1999            & 9,551\;     & 12,987\;   & 4\; & A subset of the New York Times Annotated Corpus from 1999. \\
    nytimes-2003            & 11,527\;     & 15,001\;   & 7\; & A subset of the New York Times Annotated Corpus from 2003.  \\
    wikipedia-high      & 5,738\;     & 17,311\;    & 6\; & Subset of 2014 Wikipedia dump, where articles are assigned labels based on their high level WikiProject.  \\
   wikipedia-low       & 4,986\;     & 15,441\;    & 10\; & Subset of 2014 Wikipedia dump, where articles are labeled with fine-grained WikiProject sub-groups. \\ 
   20-newsgroups       & 18,662\;     & 9,954\;    & 20\; & Collection of posts from 20 different internet newsgroups. \\ 
    20-topics       & 40,498\;     & 32,464\;    & 20\; & Tweets from user accounts associated with 20 different topical areas.\\ \hline
\end{tabular}
\label{tab:datasets}
\end{center}
\end{table*}

In this section we comprehensively assess the problem of instability in topic modeling for standard NMF and LDA approaches on a diverse collection of corpora, and examine the extent to which superior initialization and ensemble methods can improve the stability of NMF-based approaches, while also yielding accurate models.


\label{sec:datasets}

\subsection{Datasets}
	
For our experiments, we use a diverse set of ten corpora, including both high-quality long texts and user-generated content. All of these corpora have human annotated ``ground truth'' topical categories, allowing us to evaluate model accuracy. Six of these datasets consist of news articles from individual mainstream news sources (BBC, The New York Times, The Guardian, and The Irish Times), categorized by subject matter. Two more datasets consist of pages from a 2014 Wikipedia dump, categorized by their associated WikiProject. These eight datasets were previously used for topic modeling evaluations \citep{greene14topics}. We also include the popular \emph{20-newsgroups} dataset, where the ground truth categories correspond to individual newsgroups (\eg ``comp.graphics'', ``comp.windows.x'',``rec.autos'').

To evaluate performance on social media data, we include a newly-collected corpus in our experiments, known as the \emph{20-topics} dataset, which consists of 4,170,382 tweets from 1,200 prominent Twitter accounts. These accounts have been manually assigned to 20 different categories (\eg ``aviation'', ``health'', ``tech''). Each document in the corpus corresponds to the concatenation of the tweets posted by a single user for a given week during the period March 2015 to February 2016. The corpus contains 40,498 such ``user documents''. A detailed summary of all corpora used in our experiments is provided in \reftab {tab:datasets}.

\subsection{Experimental Setup}
When pre-processing the corpora, terms appearing in $< 20$ documents are filtered. We use a single list of common English stop-words for all datasets. LDA operates on bag-of-words text representations, and so was applied to the raw frequency values. For NMF, the same documents were transformed to log-based Term Frequency-Inverse Document Frequency (TF-IDF) vectors, and document length normalization was subsequently applied to produce the final document-term matrix.

In our experiments, we compare five different topic modeling approaches:
\begin{enumerate}
\item Standard LDA with random seeding, using the popular \emph{Mallet} implementation with Gibbs sampling \citep{mccallum02mallet}.
\item NMF with random initialization, using the fast alternating least squares variant proposed by \citep{lin07nmf} and provided by the \emph{sckit-learn} toolkit \citep{pedregosa11scikit}. 
\item NMF with non-random NNDSVD initialization, also implemented in \emph{sckit-learn}.
\item Basic ensemble topic modeling for matrix factorization with random initialization, as described in \refsec{sec:ensemble}.
\item K-Fold ensemble topic modeling for matrix factorization combined with improved initialization, as described in \refsec{sec:kfold}.
\end{enumerate}
For these approaches, there are a number of common and distinct parameters which need to be specified:
\begin{description}
\item[Common parameters:]
For all approaches, the number of topics \emph{k} is set to correspond to the number of ground truth categories for each dataset. 
\item[NMF parameters:]
For NMF with both random and NNDSVD initialization, the maximum number of iterations is set to 100 by default. For the random case, a different random seed is used for each run to populate values in the initial factors $\m{W}$ and $\m{H}$. This process is repeated for $r=100$ runs.
\item[LDA parameters:]
The LDA algorithm has two additional hyperparameters. We use the Mallet default values, with $\alpha=5$ and $\beta=0.01$. The maximum number of iterations is set to 1000. For each run, a different random seed is used to initialize the Gibbs sampling process. This process is repeated for $r=100$ runs.
\item[Basic ensemble:]
For the first ensemble approach, we integrate a collection of 100 members, generated via random initialization. The final number of topics $k'$ is set to be the same as the number of ground truth categories for each dataset.  This entire process is repeated 20 times to allow us to assess stability.
\item[K-Fold ensemble:]
For the second ensemble approach, we apply $p=10$ rounds of $f=10$ folds, thus also yielding a collection of 100 ensembles members for integration, with $k'$ determined as above. Again this entire process is repeated 20 times.
\end{description}


\subsection{Model Stability}
 
To assess the stability of a collection of models generated by each algorithm, we use the term-based measures \textbf{ADSD} (\reft{eqn:adsd}) and \textbf{ATS} (\reft{eqn:ats}) using the top $t=10$ terms for each topic, and the document-level \textbf{PNMI} measure (\reft{eqn:pnmi}). Results for these measures are shown in Tables \ref{tab:termdiff}, \ref{tab:termstability}, and \ref{tab:docstability} respectively. Across all three measures, we observe that the NNDSVD NMF and K-Fold approaches clearly yield the most stable results. Both of these methods produce models with perfect stability for the majority of our datasets - \ie they yield models in which the topic-term and document-topic associations remain the same. As expected, the randomly-initialized approaches perform the worst due to their inherent instability caused by stochastic elements as can be identified by their standard deviation scores. While the basic ensemble approach yields high stability for the smaller corpora, we do see some variation between different runs at the term and document level for the larger corpora. Here, the random initialization used when generating the ensemble members is still leading to variation in the results at the final ensemble integration phase, even with 100 ensemble members. However, the more structured nature of the generation phase for the K-Fold approach effectively negates this problem.

It is interesting to observe that, for the \emph{20-newsgroups} and \emph{20-topics} datasets, which contain noisier user-generated content and a larger number of underlying topics, the K-Fold ensemble approach yields higher levels of stability than NNDSVD-initialized NMF. This suggests that combining the subsampling element of the ensemble process with a structured NNDSVD initialization produces a more reliable solution. 

It is important to note that the widely-used implementation of NNDSVD provided by the \emph{sckit-learn} toolkit, as used in our experiments, relies on an approximate truncated singular value decomposition method involving randomization, in order to make it applicable to large data matrices. While the resulting decompositions are often identical, this is not always the case. Computing a full SVD would eliminate the instability, with the trade-off that the running time requirements for decomposing a large, high-dimensional document-term matrix would increase dramatically.

￿
\begin{table*}[!t]
\begin{center}
\caption{Model stability --- Comparison of Average Descriptor Set Difference (ADSD) scores for five topic modeling approaches.}
\vskip -0.5em
\begin{tabular}{|p{2.5cm}|P{2cm}P{2cm}P{2cm}P{2cm}P{2cm}|}
\hline
\textbf{Corpus}  & \textbf{LDA}  & \textbf{NMF}     & \textbf{NNDSVD}  & \textbf{Ensemble}    & \textbf{K-Fold} \\
\hline
 bbc             & 0.14 $\pm$ 0.15 & 0.15 $\pm$ 0.25 & \B{\B{0.00 $\pm$ 0.00}} & \B{0.00 $\pm$ 0.00} & \B{0.00 $\pm$ 0.00} \\
 bbc-sport       & 0.29 $\pm$ 0.14 & 0.21 $\pm$ 0.21 & \B{0.00 $\pm$ 0.00} & \B{0.00 $\pm$ 0.00} & \B{0.00 $\pm$ 0.00} \\
 guardian-2013   & 0.15 $\pm$ 0.16 & 0.18 $\pm$ 0.20 & \B{0.00 $\pm$ 0.00} & \B{0.00 $\pm$ 0.00} & \B{0.00 $\pm$ 0.00} \\
 irishtimes-2013 & 0.35 $\pm$ 0.13 & 0.42 $\pm$ 0.22 & \B{0.00 $\pm$ 0.00} & 0.01 $\pm$ 0.01 & 0.01 $\pm$ 0.01 \\
 nytimes-1999    & 0.23 $\pm$ 0.18 & 0.36 $\pm$ 0.22 & \B{0.00 $\pm$ 0.00} & 0.21 $\pm$ 0.23 & \B{0.00 $\pm$ 0.00} \\
 nytimes-2003    & 0.47 $\pm$ 0.13 & 0.31 $\pm$ 0.18 & \B{0.00 $\pm$ 0.00} & 0.01 $\pm$ 0.01 & \B{0.00 $\pm$ 0.00} \\
 wikipedia-high  & 0.26 $\pm$ 0.12 & 0.21 $\pm$ 0.13 & \B{0.00 $\pm$ 0.00} & 0.01 $\pm$ 0.01 & \B{0.00 $\pm$ 0.00} \\
 wikipedia-low   & 0.21 $\pm$ 0.06 & 0.12 $\pm$ 0.09 & \B{0.00 $\pm$ 0.00} & 0.06 $\pm$ 0.07 & 0.00 $\pm$ 0.01 \\
 20-newsgroups   & 0.31 $\pm$ 0.06 & 0.32 $\pm$ 0.09 & 0.18 $\pm$ 0.08 & 0.12 $\pm$ 0.05 & \B{0.05 $\pm$ 0.04} \\
 20-topics       & 0.23 $\pm$ 0.06 & 0.23 $\pm$ 0.09 & 0.07 $\pm$ 0.07 & 0.11 $\pm$ 0.06 & \B{0.01 $\pm$ 0.01} \\
\hline
\end{tabular}
\label{tab:termdiff}
\end{center}
\end{table*}

￿
\begin{table*}[!t]
\begin{center}
\caption{Model stability --- Comparison of term stability scores for five topic modeling approaches, calculated using Average Term Stability (ATS).}
\vskip -0.5em
\begin{tabular}{|p{2.5cm}|P{2cm}P{2cm}P{2cm}P{2cm}P{2cm}|}
\hline
\textbf{Corpus}  & \textbf{LDA}  & \textbf{NMF}     & \textbf{NNDSVD}  & \textbf{Ensemble}    & \textbf{K-Fold} \\
\hline
 bbc             & 0.86 $\pm$ 0.14 & 0.88 $\pm$ 0.21 & \B{1.00 $\pm$ 0.00} & \B{1.00 $\pm$ 0.00} & \B{1.00 $\pm$ 0.00} \\
 bbc-sport       & 0.68 $\pm$ 0.15 & 0.80 $\pm$ 0.20 & \B{1.00 $\pm$ 0.00} & \B{1.00 $\pm$ 0.00} & \B{1.00 $\pm$ 0.00} \\
 guardian-2013   & 0.83 $\pm$ 0.18 & 0.83 $\pm$ 0.19 & \B{1.00 $\pm$ 0.00} & \B{1.00 $\pm$ 0.00} & \B{1.00 $\pm$ 0.00} \\
 irishtimes-2013 & 0.61 $\pm$ 0.14 & 0.64 $\pm$ 0.18 & \B{1.00 $\pm$ 0.00} & 0.99 $\pm$ 0.01 & 0.99 $\pm$ 0.01 \\
 nytimes-1999    & 0.65 $\pm$ 0.19 & 0.72 $\pm$ 0.17 & \B{1.00 $\pm$ 0.00} & 0.82 $\pm$ 0.19 & \B{1.00 $\pm$ 0.00} \\
 nytimes-2003    & 0.53 $\pm$ 0.12 & 0.76 $\pm$ 0.14 & \B{1.00 $\pm$ 0.00} & 0.99 $\pm$ 0.01 & \B{1.00 $\pm$ 0.00} \\
 wikipedia-high  & 0.79 $\pm$ 0.15 & 0.77 $\pm$ 0.13 & \B{1.00 $\pm$ 0.00} & 0.99 $\pm$ 0.01 & \B{1.00 $\pm$ 0.00} \\
 wikipedia-low   & 0.79 $\pm$ 0.10 & 0.88 $\pm$ 0.08 & \B{1.00 $\pm$ 0.00} & 0.94 $\pm$ 0.07 & 0.99 $\pm$ 0.01 \\
 20-newsgroups   & 0.65 $\pm$ 0.07 & 0.66 $\pm$ 0.09 & 0.81 $\pm$ 0.08 & 0.86 $\pm$ 0.05 & \B{0.93 $\pm$ 0.05} \\
 20-topics       & 0.69 $\pm$ 0.08 & 0.78 $\pm$ 0.08 & 0.94 $\pm$ 0.07 & 0.89 $\pm$ 0.06 & \B{0.99 $\pm$ 0.01} \\
\hline
\end{tabular}
\label{tab:termstability}
\end{center}
\end{table*}

￿
\begin{table*}[!t]
\begin{center}
\caption{Model stability --- Comparison of document stability scores for five topic modeling approaches, based on Pairwise Normalized Mutual Information (PNMI).}
\vskip -0.5em
\begin{tabular}{|p{2.5cm}|P{2cm}P{2cm}P{2cm}P{2cm}P{2cm}|}
\hline
\textbf{Corpus}  & \textbf{LDA}  & \textbf{NMF}     & \textbf{NNDSVD}  & \textbf{Ensemble}    & \textbf{K-Fold} \\
\hline
 bbc             & 0.86 $\pm$ 0.09 & 0.89 $\pm$ 0.10 & \B{1.00 $\pm$ 0.00} & \B{1.00 $\pm$ 0.00} & \B{1.00 $\pm$ 0.00} \\
 bbc-sport       & 0.74 $\pm$ 0.08 & 0.87 $\pm$ 0.09 & \B{1.00 $\pm$ 0.00} & \B{1.00 $\pm$ 0.00} & \B{1.00 $\pm$ 0.00} \\
 guardian-2013   & 0.84 $\pm$ 0.07 & 0.88 $\pm$ 0.07 & \B{1.00 $\pm$ 0.00} & \B{1.00 $\pm$ 0.00} & \B{1.00 $\pm$ 0.00} \\
 irishtimes-2013 & 0.73 $\pm$ 0.07 & 0.81 $\pm$ 0.07 & \B{1.00 $\pm$ 0.00} & 0.99 $\pm$ 0.00 & \B{1.00 $\pm$ 0.00} \\
 nytimes-1999    & 0.69 $\pm$ 0.09 & 0.67 $\pm$ 0.14 & \B{1.00 $\pm$ 0.00} & 0.83 $\pm$ 0.13 & 0.98 $\pm$ 0.01 \\
 nytimes-2003    & 0.66 $\pm$ 0.06 & 0.78 $\pm$ 0.07 & \B{1.00 $\pm$ 0.00} & 0.97 $\pm$ 0.01 & 0.99 $\pm$ 0.00 \\
 wikipedia-high  & 0.86 $\pm$ 0.07 & 0.86 $\pm$ 0.04 & \B{1.00 $\pm$ 0.00} & 0.99 $\pm$ 0.00 & \B{1.00 $\pm$ 0.00} \\
 wikipedia-low   & 0.89 $\pm$ 0.04 & 0.89 $\pm$ 0.03 & \B{1.00 $\pm$ 0.00} & 0.96 $\pm$ 0.04 & \B{1.00 $\pm$ 0.00} \\
 20-newsgroups   & 0.63 $\pm$ 0.02 & 0.69 $\pm$ 0.04 & 0.80 $\pm$ 0.05 & 0.89 $\pm$ 0.04 & \B{0.96 $\pm$ 0.03} \\
 20-topics       & 0.84 $\pm$ 0.03 & 0.84 $\pm$ 0.03 & 0.97 $\pm$ 0.03 & 0.97 $\pm$ 0.02 & \B{1.00 $\pm$ 0.00} \\
\hline
\end{tabular}
\label{tab:docstability}
\end{center}
\end{table*}

\subsection{Model Quality}

The primary focus of our work is on model stability. But as noted in \refsec{sec:redundant}, stability without meaningful and coherent topics is unlikely to be useful. Therefore we consider the quality of the models produced by each of the five methods, in terms of accuracy and coherence.

\begin{description}
\item[Partition accuracy:] In either the case of NMF or LDA, we can convert a model's document-topic associations into a disjoint partition. We can then compare this partition with the corresponding disjoint ``ground truth'' categories for each corpus using NMI \citep{strehl02multiple}.
\item[Topic coherence:] 
Coherence refers to the overall quality and the semantic relatedness of the terms appearing in a topic descriptor. While a range of measures have been proposed in the literature, we employ a widely-used measure, Normalized Pointwise Mutual Information (NPMI) \citep{BoumaNPMI2009}, which uses term co-occurrence counts from the full corpus to measure the average coherence of the topics in a given model, based on the top $t$ terms in their descriptors. In our evaluations we use $t=10$ terms.
\end{description}

With regards to the NPMI coherence of the topics produced, \reftab{tab:coherence} shows that our proposed basic ensemble and K-Fold approaches perform the best. However, it should be noted that in most cases the differences in the average coherence scores are small. The most noticeable gap is between the LDA approach and the other NMF-based approaches, which may reflect the tendency of LDA to produce more generic and less semantically-coherent terms \citep{ocallaghan15eswa}. 

￿
\begin{table*}[!t]
\begin{center}
\caption{Model quality --- Comparison of topic coherence scores for five topic modeling approaches, based on Normalized Pointwise Mutual Information (NPMI).}
\vskip -0.5em
\begin{tabular}{|p{2.5cm}|P{2cm}P{2cm}P{2cm}P{2cm}P{2cm}|}
\hline
\textbf{Corpus}  & \textbf{LDA}  & \textbf{NMF}     & \textbf{NNDSVD}  & \textbf{Ensemble}    & \textbf{K-Fold} \\
\hline
 bbc             & 0.09 $\pm$ 0.01 & 0.15 $\pm$ 0.01 & \B{0.16 $\pm$ 0.00} & \B{0.16 $\pm$ 0.00} & \B{0.16 $\pm$ 0.00} \\
 bbc-sport       & 0.11 $\pm$ 0.01 & 0.14 $\pm$ 0.01 & 0.14 $\pm$ 0.00 & \B{0.15 $\pm$ 0.00} & 0.14 $\pm$ 0.00 \\
 guardian-2013   & 0.11 $\pm$ 0.01 & 0.15 $\pm$ 0.01 & \B{0.16 $\pm$ 0.00} & \B{0.16 $\pm$ 0.00} & \B{0.16 $\pm$ 0.00} \\
 irishtimes-2013 & 0.08 $\pm$ 0.01 & 0.12 $\pm$ 0.01 & \B{0.13 $\pm$ 0.00} & \B{0.13 $\pm$ 0.00} & \B{0.13 $\pm$ 0.00} \\
 nytimes-1999    & 0.10 $\pm$ 0.01 & \B{0.12 $\pm$ 0.01} & \B{0.12 $\pm$ 0.01} & \B{0.12 $\pm$ 0.01} & \B{0.12 $\pm$ 0.01} \\
 nytimes-2003    & 0.10 $\pm$ 0.03 & \B{0.12 $\pm$ 0.01} & \B{0.12 $\pm$ 0.01} & 0.10 $\pm$ 0.02 & \B{0.12 $\pm$ 0.01} \\
 wikipedia-high  & 0.15 $\pm$ 0.01 & 0.23 $\pm$ 0.01 & \B{0.24 $\pm$ 0.00} & 0.23 $\pm$ 0.00 & \B{0.24 $\pm$ 0.00} \\
 wikipedia-low   & 0.19 $\pm$ 0.01 & 0.24 $\pm$ 0.01 & \B{0.25 $\pm$ 0.00} & 0.24 $\pm$ 0.00 & \B{0.25 $\pm$ 0.00} \\
 20-newsgroups   & 0.12 $\pm$ 0.01 & 0.15 $\pm$ 0.01 & \B{0.16 $\pm$ 0.01} & 0.15 $\pm$ 0.00 & \B{0.16 $\pm$ 0.00} \\
 20-topics       & 0.19 $\pm$ 0.01 & \B{0.30 $\pm$ 0.01} & \B{0.30 $\pm$ 0.00} & 0.29 $\pm$ 0.00 & 0.29 $\pm$ 0.00 \\
\hline
\end{tabular}
\label{tab:coherence}
\end{center}
\end{table*}

￿
\begin{table*}[!t]
\begin{center}
\caption{Model quality --- Comparison of document partition accuracy scores for five topic modeling approaches, based on Normalized Mutual Information (NMI).}
\vskip -0.5em
\begin{tabular}{|p{2.5cm}|P{2cm}P{2cm}P{2cm}P{2cm}P{2cm}|}
\hline
\textbf{Corpus}  & \textbf{LDA}  & \textbf{NMF}     & \textbf{NNDSVD}  & \textbf{Ensemble}    & \textbf{K-Fold} \\
\hline
 bbc            & 0.81 $\pm$ 0.06 & 0.79 $\pm$ 0.04 & \B{0.82 $\pm$ 0.00} & 0.79 $\pm$ 0.00 & 0.80 $\pm$ 0.00 \\
 bbc-sport       & 0.68 $\pm$ 0.05 & 0.80 $\pm$ 0.06 & 0.83 $\pm$ 0.00 & \B{0.85 $\pm$ 0.00} & \B{0.85 $\pm$ 0.00} \\
 guardian-2013   & 0.77 $\pm$ 0.04 & 0.82 $\pm$ 0.04 & 0.83 $\pm$ 0.00 & \B{0.84 $\pm$ 0.00} & \B{0.84 $\pm$ 0.00} \\
 irishtimes-2013 & 0.68 $\pm$ 0.04 & 0.72 $\pm$ 0.04 & \B{0.77 $\pm$ 0.00} & 0.76 $\pm$ 0.00 & \B{0.77 $\pm$ 0.00} \\
 nytimes-1999    & \B{0.64 $\pm$ 0.04} & 0.53 $\pm$ 0.05 & 0.49 $\pm$ 0.00 & 0.51 $\pm$ 0.02 & 0.51 $\pm$ 0.00 \\
 nytimes-2003    & \B{0.60 $\pm$ 0.03} & 0.58 $\pm$ 0.04 & 0.55 $\pm$ 0.00 & 0.56 $\pm$ 0.01 & 0.58 $\pm$ 0.00 \\
wikipedia-high       & 0.69 $\pm$ 0.02 & 0.72 $\pm$ 0.01 & 0.72 $\pm$ 0.00 & 0.73 $\pm$ 0.00 & \B{0.74 $\pm$ 0.00} \\
wikipedia-low        & 0.84 $\pm$ 0.02 & 0.87 $\pm$ 0.03 & 0.86 $\pm$ 0.00 & \B{0.89 $\pm$ 0.02} & 0.88 $\pm$ 0.00 \\
20-newsgroups           & 0.48 $\pm$ 0.01 & 0.47 $\pm$ 0.02 & \B{0.49 $\pm$ 0.01} & 0.48 $\pm$ 0.01 & 0.48 $\pm$ 0.01 \\
20-topics       & 0.84 $\pm$ 0.02 & 0.83 $\pm$ 0.02 & 0.85 $\pm$ 0.01 & 0.87 $\pm$ 0.01 & \B{0.88 $\pm$ 0.00} \\
\hline
\end{tabular}
\label{tab:accuracy}
\end{center}
\end{table*}


To provide a clearer measure of model quality, we next consider the quality of the five methods by evaluating the partition accuracy with respect to the ground truth labels. \reftab{tab:accuracy} shows the means and standard deviations of the NMI scores for the methods on the ten corpora. Here we see that the best-performing algorithms are NNDSVD-initialized NMF and the K-Fold ensemble approach, although this varies with the dataset. The randomly-initialized algorithms exhibit considerable variation in the quality of the models they produce, as indicated by the standard deviation scores, and are worse on average than the ensemble and SVD-based methods, with the exception of the case where LDA is applied to the two New York Times corpora.

\subsection{Statistical Significance}

To further investigate the differences between the algorithms, we performed a series of statistical tests on the results presented in the previous section. We carried out a non-parametric Friedman's Aligned Rank test \citep{garcia2010advanced} for each of the five measures previously reported (ADSD, ATS, PNMI, NPMI, and NMI) to test for the presence of statistically significant differences in the results amongst the five algorithms and across the ten datasets. These tests returned p-values of 0.000004 (ADSD), 0.000002 (ATS), 0.000003 (PNMI), 0.00002 (NPMI), and 0.118 (NMI) respectively. This indicates that statistically significant differences, at the 1\% confidence level, exist in the results achieved by the different algorithms for each measure, except for partition accuracy (NMI).

To determine if there was a statistically significant difference between our proposed K-Fold algorithm and the other topic modeling approaches with respect to the four remaining measures, we performed a series of Friedman's Aligned Rank Pairwise post hoc tests \citep{garcia2010advanced}, with the K-Fold approach used as a control. The results from these tests are reported in \reftab{tab:significance}. It is interesting that there is a statistically significant difference, at the 1\% confidence level, between our proposed approach and the two randomly initialized topic modeling algorithms across all measures, which along with the previously reported performance of the algorithm suggests that the K-Fold approach produces more stable and higher quality topic models. There is no statistical difference between our proposed K-Fold approach, the ensemble and NNDSVD, which may indicate that these three measures are similar due to producing more deterministic solutions and this notion is further strengthened due to their similar performance regarding the measures previously reported. It is also interesting to note that there is a statistical difference between the LDA and the K-Fold approach with regards to coherence, likely due to LDA based approaches generating topics with a lower coherence \citep{ocallaghan15eswa}.



￿
\begin{table*}[!h]
\begin{center}
\caption{Posthoc p-values based on Friedman's Aligned Rank Pairwise test for each of the four measures that were statistically significant, while using the K-Fold approach as a control algorithm. * p $\leq$ 0.05, ** p $\leq$ 0.01, *** p $\leq$ 0.001, **** p $\leq$ 0.0001}
\vskip -0.5em
\begin{tabular}{|p{2cm}|llllP{2cm}|}
\hline
\textbf{Measure}  & \textbf{LDA}  & \textbf{NMF}     & \textbf{NNDSVD}  & \textbf{Ensemble}    & \textbf{K-Fold} \\
\hline
ADSD            & **** 0.00001 & **** 0.00002 & 0.645 & 0.273 & NA \\
ATS               & **** 0.000002 & **** 0.0001 & 0.713 & 0.250 & NA \\
PNMI             & **** 0.000002 & **** 0.00007 & 0.576 & 0.304 & NA \\
NPMI             & **** 0.00001 & 0.111 & 0.921 & 0.452 & NA \\
\hline
\end{tabular}
\label{tab:significance}
\end{center}
\end{table*}

￿
\begin{table*}[!t]
\begin{center}
\caption{Comparison of average running time in seconds for the five topic modeling approaches.}
\vskip -0.5em
\begin{tabular}{|p{2.5cm}|P{2.7cm}P{2.5cm}P{2.5cm}P{2.5cm}P{2.6cm}|}
\hline
\textbf{Corpus}  & \textbf{LDA}  & \textbf{NMF}     & \textbf{NNDSVD}  & \textbf{Ensemble}   & \textbf{K-Fold} \\
\hline
 bbc                     & 40.74  & \B{0.22} & 0.30 & 22.68 & 30.00\\
 bbc-sport            & 33.34 & \B{0.08} & \B{0.08} & 7.90 & 8.56 \\
 guardian-2013    & 231.33  & 1.98 & \B{1.94} & 198.89 & 189.87 \\
 irishtimes-2013  & 82.75 & \B{0.85} & 0.86 & 86.76 & 84.69 \\ 
 nytimes-1999     & 370.85 & \B{2.95} & 3.74 & 296.18 & 375.97 \\
 nytimes-2003     & 454.36 & 6.11 & \B{4.83} & 613.00 & 538.57\\ 
 wikipedia-high    & 667.88 & 5.12 & \B{3.31} & 513.68  & 319.96 \\
 wikipedia-low     & 627.28 & 6.42  & \B{4.07} & 644.70 & 380.85 \\
 20-newsgroups & 229.29 & \B{14.16} &15.15 & 1421.44 & 1527.62 \\
 20-topics           & 1802.16  & \B{72.07} &  115.02 &  7226.22  & 6793.77 \\
\hline
\end{tabular}
\label{tab:timings}
\end{center}
\end{table*}

\subsection{Computational Expense}

While the timing of algorithms can vary depending on the hardware and implementation utilized, it is useful in this case to obtain an estimate of how much longer the proposed ensemble approaches take with respect to traditional topic modeling algorithms. Each topic modeling approach was run 100 times and the average times are reported in \reftab{tab:timings}. The experiments were carried out on a machine with 12 2.4GHz cores and 128GB of RAM. These running times are impacted due to numerous factors, including the number of documents in the corpus, the dimensionality of the corresponding document-term matrix, and the number of topics $k$ selected for the corpus. As expected, it is clear that the two ensemble approaches take considerably longer to run than the other algorithms. This is naturally due to the underlying nature of their generation step, where 100 iterations of NMF have to be generated.  While the LDA \emph{Mallet} implementation is widely used for topic modeling, we observe that it is considerably slower in the majority of our experiments, in fact it is frequently slower than our proposed K-Fold approach which utilizes a more structured but slower initialization step. 

\subsection{Discussion}

So far we have discussed the two main criteria for evaluating topic modeling algorithms separately. These are the evaluation of \emph{model quality}, by examining topic coherence and partition accuracy, and the evaluation of \emph{model stability} by examining term stability and document stability. However, it is important to note that the output of both criteria should be considered together -- our evaluations highlight that some topic modeling approaches perform well with respect to one criterion, while performing poorly with respect to the other. 
An example of this can be seen in the results produced by NNDSVD for the \emph{nytimes-1999} dataset. For the term and document stability, perfect stability scores are achieved (Tables \ref{tab:termstability} and \ref{tab:docstability}). However, when we take into account partition accuracy for the same dataset (\reftab{tab:accuracy}), the quality of this solution is not as good as it initially appears. The NNDSVD initialization actually performs the worst with regards to accuracy in this case, with a low NMI score of 0.49. Similarly, while randomly-initialized LDA out-performs all other approaches on both New York times corpora in terms of partition accuracy (\reftab{tab:accuracy}), the corresponding stability scores are consistently poor across all measures (Tables \ref{tab:termdiff}--\ref{tab:docstability}). Following the discussion of ``redundant stability'' in \refsec{sec:redundant}, these results raise an interesting problem in that, while we strive to produce the most stable results as possible, it may also be the case that these results are of poor quality from a model quality standpoint.


Among the two newly-proposed approaches, it is interesting to observe that the basic ensemble approach does not perform as well as the K-Fold approach, even though they are based on a similar ensemble process. In the case of the former, we aim to promote diversity when generating our base ensemble members by using randomly-initialized NMF, as motivated by previous work in both supervised and unsupervised ensemble learning \citep{brown05diversity,kuncheva04using}. However, for larger datasets, the stochastic nature of this approach tends to cause the final results to contain a degree of variance across different runs of the overall ensemble. In contrast, by combining structured document subsampling with NNDSVD initialization to generate each ensemble member, the K-Fold approach exhibits very little instability across the 20 runs in our experiments, as indicated in the results in Tables \ref{tab:termdiff}--\ref{tab:docstability}. These findings correspond to those of \citet{hadjitodorov06moderate}, who demonstrated that a moderate level of diversity leads to useful ensembles in cluster analysis.

Overall, among the techniques considered in our experiments, the K-Fold ensemble approach produces the best models when taking into account both quality and stability. While the observed NMI scores were lower for certain datasets, it still performs better than the other methods with respect to half of our corpora. This strategy also appears to handle noisy user-generated data well, in comparison to alternative techniques.


￿
\section{Conclusions}
\label{sec:conc}

While topic modeling methods such as LDA and NMF are widely applied in a range of domains to analyze unstructured text, researchers often do not consider the effect that random initialization has on models produced by these methods. In this paper we have demonstrated that, for both methods, this can result in significant variations in the topics produced over multiple runs over the same corpus. This effect is manifested both at the term and document level, which can potentially lead to different human interpretations of the underlying thematic structure of the data.

To address the issue of instability in the context of NMF, we have investigated the extent to which improved algorithm initialization and ensemble strategies can produce more stable models, which are almost potentially more accurate and insightful. We compared the performance of these approaches with regards to five different metrics that measure stability, accuracy, and coherence of topics. Our results indicate that a new K-Fold ensemble approach afforded the most stable and accurate set of models, although initializating NMF based on a SVD approximation of the document-term matrix can also provide a clear improvement over standard NMF and LDA methods.

One concern that arises in the application of ensemble learning techniques in general relates to scalability. While the ensemble techniques described in this paper can be naturally parallelized, there is considerable scope for reducing the computation time required to generate the ensemble. A potentially promising idea to investigate in this context relates to the concept of \emph{snapshot ensembles} from supervised learning \citep{huang17snapshot}, where a single algorithm run is allowed to converge to several local minima during the optimization process, each providing a contribution to the overall ensemble. Such an approach might also be used to yield more stable topic models via matrix factorization and reduce the computational expense.

\vspace{3mm}
\noindent\textbf{Acknowledgement.}  This research was supported by Science Foundation Ireland (SFI) under Grant Number SFI/12/RC/2289.

\bibliographystyle{modelsarticle-harv}
\bibliography{topic-stability} 

\end{document}